\documentclass[12pt,oneside,english]{article}

 \usepackage{epsf}
\usepackage[T1]{fontenc}
\usepackage[latin1]{inputenc}
\usepackage{babel}
\usepackage{setspace}
\makeatletter
\begin{document}
\section*{\bf Phase transitions  in a one-dimensional multibarrier potential 
of finite range  } 
\makeatother
\bigskip \bigskip 
\centerline{\large D.Bar$^{a}$ and L.P.Horwitz$^{a,b}$ } 
{\bf $^a$Department of Physics, Bar Ilan University, Ramat Gan,
Israel \\ $^b$Raymond and Beverly Sackler Faculty of Exact Science, School of
Physics, Tel Aviv University, Ramat Aviv, Israel}

\bigskip \bigskip

\begin{abstract}   \noindent 
{\it  
We have previously studied properties of a one-dimensional potential with $N$ equally
spaced identical barriers in a (fixed) finite interval for both finite and
infinite $N$. It was observed that scattering and spectral properties depend
sensitively on the ratio $c$ of spacing to width of the barriers (even in the
limit $N \to \infty$). We compute here the specific heat of an ensemble of such
systems and show that there is critical dependence on this parameter, as well as
on the temperature, strongly suggestive of phase transitions. }
\end{abstract}
\noindent
{PACS number(s): 05.70.Fh, 02.60.Cb, 03.65.Ge }
\bigskip \noindent 
\protect \section{Introduction \label{sec1}}
We have studied the one-dimensional locally periodic multibarrier potential of 
finite range for both finite and infinite number $N$ of barriers. 
We  found \cite{Bar1} that there is a critical dependence  of the transmission
coefficient, the cross section  and the 
distribution of poles of the $S$-matrix \cite{Merzbacher} 
 on the ratio $c$  of the total interval between the barriers to
their total width.  Under certain conditions this model was found to contain 
signatures of chaotic behaviour \cite{Reichl}, such as rapid spread of wave
packets and Wigner type spectral characteristics. \par
We discuss in this work thermodynamic properties  
 \cite{Reif} such as the specific heat and the entropy of an ensemble of such 
 systems  and
show that they also depend critically upon this ratio in addition to  their dependence 
upon the temperature. We show that when the number of barriers is not large 
there are no low-lying energy eigenvalues  for small
values of the ratio $c$. These  values of $c$ depend upon the value of 
the number $N$ of 
barriers in such a manner that the ranges  of $c$ in which this part of the
energy spectrum is missing increase at first with $N$ and then decrease as $N$ 
continue
to grow  until they disappear entirely for large enough
$N$. Above the upper  boundaries  of  these ranges are the points where the 
energy
spectrum contains eigenvalue distribution over all energies.  A similar
phenomenon occurs in the work of Fendley and Tchernyshyov \cite{Fendley} on
one-dimensional phase transitions, where the seriously disordered phase behaves
as if it were at very high temperature, where the effect of interactions is
relatively small, i.e, as an almost independent particle model.  This apparently 
accounts, even
for neighbouring values of $c$ within these ranges,   
 for  large changes in
the average energy and all the other statistical mechanics properties, such as
specific heat and entropy,  derived
from it.  For example, the corresponding curves of the specific heat, for these
neighbouring values of $c$, as functions of the temperature, differ markedly
from each other,  as will be shown in the following, when plotting these curves
in a single figure. This sensitivity of the specific heat is largest for
intermediate values of $N$ and $c$ in which the curves of the specific heats as
functions of the temperature differ to the extent that some of them exhibit
double peak phase transitions while other curves, for neighbouring values of $c$,
resemble the conventional Debye curve \cite{Reif}  which characterizes the solid
crystal. For infinite $N$, as will be shown, the double peak phase phenomenon is
retained even for large values of $c$.    Also, we find, for all values of $N$ and $c$, 
 indication of phase transitions in specific heat  at small values of the temperature
$T$. We note that for any $c$, which is  greater than some value which 
depends upon $N$, the
system behaves for all $N$, except for a spike form at small $T$,  in a manner
similar to that of the solid crystals in which the constituent atoms are widely
separated, and thus, as we have remarked, are characterized by large values of the
ratio $c$. This is seen on the corresponding graphs of  the specific heats as functions of the temperature which have the
same form as that of Debye (we do not imply here that our system is analogous to
a system of lattice vibrations, but only to the behaviour of the result due to
 the apparent presence of uncorrelated modes).    
\par 
In Section 2 we study the properties of a one-dimensional $N$ potential 
barrier system when $N$ is a finite number. In Section 3 we discuss  the limit 
of $N \to \infty$ (in the same fixed interval).  For both cases we find 
 abrupt and large changes in the values of specific heat and entropy for small
 values of the temperature $T$.  
  These  large changes suggest  the existence of 
phase transitions.  In particular, for intermediate $N$ and $c$, 
   we find double peaks \cite{Leung,Tanaka,Kim,Ko} in the specific heat curves. 
   Double peak phase transitions appear, especially, for  infinite $N$ in
   which  it remains effective even for large values of $c$. Tanaka et al 
\cite{Tanaka} have found double peak structures in anti-ferromagnetic materials, corresponding
to magnetic phases, where the external magnetic field seems to play a role
somewhat analogous to the parameter $c$ in our study. Leung and Neda 
\cite{Leung} and also 
Kim et al \cite{Kim} have found
double peaks in the response curves, apparently associated with dynamically 
induced phase transitions. Ko and Asakawa \cite{Ko} have also found a double
peak structure in their calculations of the phases of a quark-gluon plasma, 
where one may think of a large number of interactions in a bounded region.   We
do not imply that our model contains an analog of their interacting systems, but
suggest that some of the mathematical properties may be common.   
\noindent

\bigskip \noindent 
\protect \section{The one-dimensional $N$ potential barrier system  \label{sec2}}
We consider a  finite $N$ barrier system where all these  barriers have the 
same height $v$ and are locally periodic in the finite interval.  This
array is assumed to start at the point $x=-\frac{a+b}{2}$ 
and ends at  $x=\frac{a+b}{2}$, so that the total length of this system is $L=a+b$. 
Here $a$ is the total width of all the $N$ barriers (where $v \ne 0$), and $b$
is the total sum of all the intervals between neighbouring barriers 
(where $v = 0$). Thus, since we have $N$ potential barriers the width of each
one is  $\frac{a}{N}$, and the interval between each two neighbouring ones is 
$\frac{b}{N-1}$. Denoting $b=ac$ where $c$ is a real number we can express $a$
and $b$ in terms of $L$ and $c$ as  \cite{Bar1} 
$a=\frac{L}{1+c}, \; \; \; \; \;  b=\frac{Lc}{1+c}$. \par  
Let us first consider the passage of a plane wave through this system, which has
the form $\phi=A_1e^{ikx}+B_1e^{-ikx}, \ \ \   x \le  -\frac{a+b}{2}$. 
Matching boundary conditions at the beginning and end of each barrier,  we may 
construct a solution in terms of the transfer matrices
\cite{Merzbacher,Yu}  $P^{(j)}$ on the $j$th
barrier. After the $n$th barrier we obtain, using the terminology in
\cite{Merzbacher}, the following transfer equation  \cite{Bar1}
 \begin{equation} \label{e1} \left[ \begin{array}{c} A_{2n+1} \\ B_{2n+1} 
\end{array} \right] = P^{(n)}P^{(n-1)}\ldots P^{(2)}P^{(1)}
\left[ \begin{array}{c} A_1\\ B_1 \end{array} \right],  
\end{equation} where $p^{(n)}$  is a product of three two dimensional matrices
\cite{Bar1}. $A_{2n+1}$ and $B_{2n+1}$ are the 
amplitudes of the transmitted and reflected parts respectively of the wave function 
at the $n$th potential barrier.   $A_1$ is the coefficient of the initial wave
that approaches the potential barrier system, and $B_1$ is the coefficient of the 
reflected wave from the first barrier.   Eq (\ref{e1}) may be written as
\cite{Bar1} \begin{eqnarray}  &&\left[
\begin{array}{c} A_{2n+1} \\ B_{2n+1} 
\end{array} \right]=\left[ \begin{array}{c c} e^{-ik(a+b-\frac{a}{2N})}
 & 0 \\0&e^{ik(a+b-\frac{a}{2N})} \end{array} \right] \left[ \begin{array}{c c}
T_{11}&T_{12}
\\T_{21}&T_{22}\end{array} \right] \Biggl( \left[ \begin{array}{c c}
e^{\frac{ikb}{N-1}}&0\\
0&e^{-\frac{ikb}{N-1}} \end{array} \right] \left[ \begin{array}{c c}
T_{11}&T_{12}
\\T_{21}&T_{22}\end{array} \right] \Biggr) ^{n-1} \cdot \nonumber  \\ 
&& \cdot \left[ \begin{array}{c c}
e^{\frac{-ika}{2N}}&0\\
0&e^{\frac{ika}{2N}} \end{array} \right]\left[ \begin{array}{c} A_1 \\ B_1 
\end{array} \right] =Q \left[ \begin{array}{c} A_1 \\ B_1 
\end{array} \right],  \label{e2} 
 \end{eqnarray}
where $k$ is
  $\sqrt{\frac{2me}{\hbar^2}}$. Note that the transfer equation (\ref{e2}) is
  valid for both cases of $e>v$ and $v>e$  except for the two-dimensional matrix
  in $T$ which, for the $e>v$  case, assumes the form      
  \begin{eqnarray}  && T_{11} = 
 \cos(\frac
{aq}{N})+i\frac{\xi}{2}\sin(\frac{aq}{N}), \; \; \; T_{12} =i\frac{\eta}{2}
\sin(\frac{aq}{N}) \label{e3} \\ &&T_{21} = -i\frac{\eta}{2}
\sin(\frac{aq}{N}), \; \; \; T_{22} = 
 \cos(\frac
{aq}{N})-i\frac{\xi}{2}\sin(\frac{aq}{N}),  \nonumber  \end{eqnarray}  
and for the $v>e$ case 
\begin{eqnarray}  && T_{11} =
 \cosh(\frac
{aq}{N})+\frac{\grave \xi}{2}\sinh(\frac{aq}{N}), \; \; \;  T_{12}
=\frac{\grave \eta}{2}
\sinh(\frac{aq}{N}) \label{e4} \\ && T_{21} = -\frac{\grave \eta}{2}
\sinh(\frac{aq}{N}), \; \; \;  T_{22} = 
 \cosh(\frac
{aq}{N})-\frac{\grave \xi}{2}\sinh(\frac{aq}{N})  \nonumber  \end{eqnarray}
The $q$, $\xi$ and $\eta$ of Eq (\ref{e3}) are  \cite{Bar1} 
 $ q=\sqrt{\frac{2m(e-v)}{\hbar^2}},\ \ \ \ \xi=\frac{q}{k}+\frac{k}{q}, \quad 
  \eta=\frac{q}{k}-\frac{k}{q},  $ 
and those  of 
Eq (\ref{e4})  \cite{Bar1} 
$ q= \sqrt{\frac{2m(v-e)}{\hbar^2}}, \ \ \ \ \grave \xi=-i\eta= \frac{q}{ik}+\frac{ik}{q}, \quad 
 \grave \eta=-i\xi=\frac{q}{ik}-\frac{ik}{q}$. 
  In the numerical part of this work we
 assign $\hbar=1$, and $m=\frac{1}{2}$. 
  Note that the two  matrices $T$ from Eqs (\ref{e3})-(\ref{e4}) do not depent on
 $n$. \par
 We find,  now,  the energy spectrum of this system.  We use 
the $S$-matrix and  the periodic boundary 
conditions at the two remotely placed sides of the system. That is, we assume 
that the wave
function and its derivative at the far right end of the system, say at $x=C$ 
where $C$ is much
 larger than the size $L=a+b$ of the system, are equal to the wave function
and its derivative at the corresponding far left end of the system at $x=-C$, 
so
we obtain 
 $ A_{2N+1}e^{ikC}=A_1e^{-ikC}$,    
$B_{2N+1}e^{-ikC}=B_1e^{ikC}$.   
Thus, using the last  relation and those  between the components
of the $S$ and $Q$ matrices \cite{Merzbacher}, where  $Q$ is given by Eq
(\ref{e2}),    
$$ S_{11}=Q_{11}-
\frac{Q_{12}Q_{21}}{Q_{22}}=\frac{1}{Q_{22}}, \; \; \; \; 
S_{12}=\frac{Q_{12}}{Q_{22}}, \; \; \; \; S_{21}=-\frac{Q_{21}}{Q_{22}}, \;
\; \; \; S_{22}=\frac{1}{Q_{22}}, $$  one may write   the following 
dependence of the 
outgoing waves $A_{2N+1}$  and  $B_1$  upon the ingoing ones $A_1$ and $B_{2N+1}$
     \begin{eqnarray} &&  \left[
\begin{array}{c} A_{2N+1} \\ B_1 
\end{array} \right]=\left[ \begin{array}{c c} S_{11}&S_{12} \\ S_{21} &S_{22} 
\end{array} \right]\left[ \begin{array}{c} A_1 \\ B_{2N+1} 
\end{array} \right]  
  =e^{2ikC} \left[ \begin{array}{c c} S_{11}&S_{12} \\ S_{21} &S_{22} 
\end{array} \right]\left[ \begin{array}{c} A_{2N+1} \\ B_1 \end{array}
\right]=\label{e5} \\ && =
\frac{e^{2ikC}}{Q_{22}}\left[ \begin{array}{c c} 1&Q_{12} \\ -Q_{21} &1 
\end{array} \right]\left[ \begin{array}{c} A_{2N+1} \\ B_1
\end{array} \right] \nonumber \end{eqnarray} 
To obtain a non trivial solution for the vector $\left[
\begin{array}{c} A_{2N+1} \\ B_1 
\end{array} \right]$  one  must solve the following equation 
\begin{eqnarray} &&  \det\left[ \begin{array}{c c} 
 \frac{e^{2ikC}}{Q_{22}}-1& \frac{e^{2ikC}Q_{12}}{Q_{22}}\\
-\frac{e^{2ikC}Q_{21}}{Q_{22}}  & \frac{e^{2ikC}}{Q_{22}}-1
\end{array} \right]  = 
\frac{\cos(4kC)}{Q_{22}^2}-\frac{2\cos(2kC)}{Q_{22}}+1+ \label{e6} \\ && +
\frac{Q_{12}Q_{21}\cos(4kC)}{Q_{22}^2}+ i(\frac{\sin(4kC)}
{Q_{22}^2}-\frac{2\sin(2kC)}{Q_{22}}+
\frac{Q_{12}Q_{21}\sin(4kC)}{Q_{22}^2})=0 \nonumber 
\end{eqnarray}  
 The form of the  equations (\ref{e5})-(\ref{e6}) are the same 
  for both cases of $e>v$ and $v>e$ except for the
 two-dimensional matrix $Q$ from Eq (\ref{e2}) in which the matrix $T$ assumes
 the form (\ref{e3}) for the  $e>v$ case and the form (\ref{e4}) for $v>e$. 
  The energies, satisfying this relation,  depend, as noted,  on  $c$ and $N$ and are obtained
numerically. We show, now, that the high energy part ($e>\!>v$) of the spectrum obtained 
from Eq (\ref{e6}) depends only on $C$ and may be obtained analytically without
using numerical methods.    For
that matter we note that when the energy $e$ becomes very large we have 
$k \approx q$, and $\xi$ and $\eta$ (see the inline equations after Eq
(\ref{e4}))  obtain the values of 
2 and 0
respectively. In that case the components $T_{12}$ and $T_{21}$ of the two
dimensional matrix $T$ from Eq (\ref{e3}) become zero, and the diagonal
elements $T_{11}$ and $T_{22}$ become $e^{\frac{iqa}{N}}$ and 
$e^{-\frac{iqa}{N}}$ respectively. Thus, the two dimensional matrix $Q$ from Eq
(\ref{e2}) becomes much simplified and may be calculated analytically as 
follows (using $q \approx k$ for the very high part of the energy spectrum)
\begin{eqnarray}  && Q_{(high-energies)}=\left[ \begin{array}
{c c} e^{-ik(a+b-\frac{a}{2N})}
 & 0 \\0&e^{ik(a+b-\frac{a}{2N})} \end{array} \right] \left[ \begin{array}{c c}
e^{\frac{ika}{N}}&0
\\0& e^{-\frac{ika}{N}} \end{array} \right] \cdot \nonumber  \\ && \cdot \Biggl( \left[ \begin{array}{c c}
e^{ik(\frac{b}{N-1}+\frac{a}{N})}&0\\
0&e^{-ik(\frac{b}{N-1}+\frac{a}{N})} \end{array} \right] \Biggr) ^{n-1} 
 \left[ \begin{array}{c c}
e^{-\frac{ika}{2N}}&0\\
0&e^{\frac{ika}{2N}} \end{array} \right]=  \label{e7} \\ && = \left[ \begin{array}{c c}  
 e^{-ik(a(1-\frac{3}{2N})+b-(n-1)(\frac{b}{N-1}+\frac{a}{N})+\frac{a}{2N})} &0 \\
 0&  e^{ik(a(1-\frac{3}{2N})+b-(n-1)(\frac{b}{N-1}+\frac{a}{N})+\frac{a}{2N})} 
 \end{array} \right]= \nonumber \\ && =\left[ \begin{array}{c c}  
 e^{-ik(a+b+\frac{b}{N-1}-n(\frac{b}{N-1}+\frac{a}{N}))} &0 \\
 0&e^{ik(a+b+\frac{b}{N-1}-n(\frac{b}{N-1}+\frac{a}{N}))}   
 \end{array} \right]=\left[ \begin{array}{c c} 1 &0\\0&1 \end{array} \right],
 \nonumber  \end{eqnarray}
 for $n=N$, required for the  application of (\ref{e6}). 
Thus we see that the two dimensional matrix $Q$  becomes the
 unity matrix in the limit of large energies $e$. Substituting
 the resulting components of $Q$ ($Q_{11}=Q_{22}=1$, 
$Q_{12}=Q_{21}=0$) in Eq (\ref{e6}) we obtain \begin{equation} 
\det\left[ \begin{array}{c c} e^{2ikC}-1 &0\\0&e^{2ikC}-1 \end{array} \right] 
=(e^{2ikC}-1)^2=0 \label{e8} \end{equation} 
The last equation is  satisfied for all $k=\frac{\pi n}{C}, \; \; \;
n=0,1,2,3 \ldots$, but since we are restricted here to the high energy part of 
the energy spectrum we refer  only to the large values of $n$. \par   
As remarked,    the previous equations  may be applied 
 also for the $e<v$ case except for the correct $Q$.  
 Thus,  the energy spectrum is composed of three parts:  1) the part for the
 $e<v$
 case. 2) the part that satisfies $e_{n_0}>e>v$, where $e_{n_0}$ is some arbitrarily 
 specified large energy (these two parts are obtained numerically by solving Eq
 (\ref{e6}) for both cases of  $e>v$ and $e<v$ case with the appropriate $Q$),   
  and  3) all the high energies that
 are larger than  $e_{n_0}$ and  are 
 obtained analytically using Eqs   (\ref{e7}), (\ref{e8}) and the  
 relation  $k=\frac{\pi n}{C} \; \;
 \; \;
 n=n_0,n_0+1,n_0+2 \ldots$, where $n_0$ is some specified large integer that
 corresponds to the  energy $e_{n_0}$. Now, since the relation between the
 energy $e$ and $k$ is $k=\sqrt{e}$ (as remarked,  we assign $\hbar=1$ and
 $m=\frac{1}{2}$), the high part of the energy spectrum is
 given by $e_{high}=(\frac{\pi n}{C})^2$, where $n=n_0,n_0+1,n_0+2,\ldots$. 
 For the numerical part of the calculations we assign the following values for
 the relevant parameters: $v=60$, $C=90$, $L=20$, and $e_{n_0}=1080$. Thus, the
 $e<v$ part of the spectrum is $0<e<60$. The second part is $60<e<1080$, and the
 third part is $e>1080$. The $n_0$ that corresponds to  $e_{n_0}$ is 
 $n_0=\frac{C}{\pi}\sqrt{e_{n_0}} \approx 941$.  The assigned values for $C$ and
 $L$ yield a ratio of $\frac{C}{L}=4.5$ which ensures that the points $x=\pm C$
 are far enough from the region of the potential and so we may assume (see the
 discussion after Eq (\ref{e4})) that the values of the wave function and its
 derivative at $x=+C$ are equal to these at $x=-C$. These equalities are needed
 for obtaining  Eqs (\ref{e5})-(\ref{e6}). Thus, the results of the numerical
 simulations depend upon the ratio $\frac{C}{L}$ and not upon the specific
 values of $C$ and $L$. The somewhat higher value of the potential $v=60$ (the
 value, conventionally chosen for numerical simulations (see \cite{Maple}), is
 from the range $(2 \le v \le 8)$) is especially chosen for the $e<v$ part of
 the  simulations in order to accumulate enough data and thus to obtain a better
 statistics. Note that the simulated 
 $e_{n_0}>e>v$ part, where $e_{n_0}=1080$, yields a large amount of data so that in
 order not to remain with a comparatively small amount for the $e<v$ part we
 have chosen, as remarked, a somewhat large value of $v$.  The value of 
  $e_{n_0}=1080$ is chosen as a limit value beyond which the corresponding terms
  of the sums $\sum e(c,N)e^{-\beta e(c,N)}$ and $\sum e^{-\beta e(c,N)}$ 
  in the
  following Eq (\ref{e9}) may be approximated by their simplified analogs
  obtained using Eq (\ref{e8}).  Thus, the results are not sensitive to these
  specific values of $C$, $L$, $v$, and $e_{n_0}$.  \par
  We want to obtain a formula for  the average energy from which we may derive
most of the statistical mechanics parameters mentioned above. This average
energy depends  upon $c$, $C$, $N$ and the   temperature $T$ (the dependence upon the
temperature is through $\beta=\frac{1}{\kappa T}$, where $\kappa$ is the
Boltzman constant which is assigned, in our numerical work,  the value of unity) 
and is given by   
\begin{equation} \label{e9}
<\!e\!>(\beta,c,N)=\frac{\sum e(c,N)e^{-\beta e(c,N)}+
\sum_{941}^{\infty}(\frac{n \pi}{C})^2
e^{-\beta(\frac{\pi n}{C})^2}}{\sum e^{-\beta e(c,N)}+
\sum_{941}^{\infty}e^{-\beta(\frac{\pi n}{C})^2}},     \end{equation} 
where the first sum in the numerator and denominator  contains  the  contributions from all $e(c,N)<1080$. 
For higher values of $e$ the expressions are simpler and we take this into
account in the second sum over all integers $941 \le n $. 
  The first sum in the numerator and denominator of Eq (\ref{e9}) 
includes  the energies from the $e<v$ and the $v<e<1080$ parts of the spectrum .
  These parts are  obtained numerically 
from Eq (\ref{e6}) 
in which we substitute for the components of $Q$ from Eq (\ref{e2}), using the
$T$'s of Eq (\ref{e3}) for the $e>v$ case, and those of Eq (\ref{e4}) for  
$e<v$.  From Eq (\ref{e9}) we obtain   an average energy for each specific
triplet of values for $c$, $N$, and $\beta$, and from this average energy we 
may derive the  quantities of statistical physics. We note that the sums over the energies from the range $0.1 \le e \le 1080$ 
depend upon the parameters $N$ and $c$ whereas the sums over the higher energies
do not depend upon them (see Eqs (\ref{e7}), (\ref{e8})).  The
specific heat $C_h$ is obtained as the derivative of the average energy from Eq 
(\ref{e9}) with
respect to the temperature $T$.  \par
It is found that for large values of the temperature $T$  the
 curves of the specific heats $C_h$, for all values of $c$ and $N$, tend to the
 constant value  $C_h=0.55$. That is, for these $T$'s the curves of $C_h$ become
 as expected  a constant curve as for Dulong-Petit \cite{Reif}.    
  Also, for  small  $T$'s the curves 
 of $C_h$, for all $c$ and $N$,  rise rapidly  to their maximum values
 $C_{h_{max}}$ 
 from which
 they descend either to the asymptotic value of 0.55 for large $T$ as noted 
 or to some minimum from which they rise again to a second peak that descends 
 to the value of 0.55 for large $T$. We note that the
  $C_{h_{max}}$'s are points at which the   derivative appears to be very large 
  and are, therefore, suggestive of the existence of phase transition 
   \cite{Reichl1}.  The values of these 
$C_{h_{max}}$, however, as well as the behavior of the specific heat for
intermediate values of $T$ depend upon $c$ and $N$. It has been found that for
large  $c$  the curves of the specific heats, as functions  
 of $T$,  are    of the Debye type \cite{Reif},  
that is, the rapid approach  to maximum 
$C_{h_{max}}$ for
small $T$ and  the immediate decrease to a constant value as $T$ grows. For 
 small $c$, however, the forms of the specific heat $C_h$ for intermediate 
  $T$ depart 
 markedely from that of Debye, and the range of $T$ in  which $C_h$ is
 different depends upon $N$ in such a manner that this range
 increases as $N$ grows. For example, when $N=2$ this range is $0.4 \le T 
\le 3.8$, for $n=35$ it is $0.4 \le T \le 100$ and for $N=250$ this range increases to
$5 \le T \le 2000$. \par Figure 1 shows 38 curves of the specific heats, 
for $N=6$,  
 as functions of
the temperature in the range $0.1 \le T \le 35$. Each curve is for a different
integral value of $c$ in the range 
$2,3,\ldots,39$. The central and
dense part of the figure, where a large part of the curves have the same
form,  are those graphs that have a large $c$ and, therefore,  may represent
the solid crystals that are characterized by a periodic structure in which 
 neighbouring occupied sites are widely separated. Indeed, these curves 
 resemble, except for the sharp peaks,  that of Debye 
which represents well the solid crystal. The other curves that differ from the
central ones, and that generally have  large values for the
specific heats at small $T$,  are those that have smaller $c$ and, therefore,  
do not have the behaviour of the specific heat curves of a solid crystal. 
These curves may represent some "soft" substance
\cite{Flowers} in which the
constituent atoms or molecules are closer to one another than in the solid
crystal. These substances  have Einstein frequencies \cite{Reif,Flowers} 
smaller than those of the solid crystals by a factor of 10 to 50
\cite{Flowers} and,  therefore, are characterized by  higher values of the
specific heats.   Also, the remarked  sharp peaks 
 have a very large value for the  
derivative and   this suggests an existence of phase transitions. 
 We have, especially,   studied the immediate  neighbourhood  of the region
containing rapid variation in specific heat as a function of temperature and the
parameter $c$ (i.e, the neighbourhood of a phase transition). We see that there
is a very strong and critical dependence on $c$. This result illustrates the
physical mechanism for the formation of such rapid transitions, a resonance-like
phenomena controlled by the {\it geometry} of the barrier system. This is 
demonstrated
 in Figure 2 which shows 15 curves of the specific heat,  for $N=6$,  as functions of the
 temperature $T$ and for the following values of $c=0.3, 0.4, 0.5, \ldots 1.7$. 
  The central
 dense part of the figure, which is composed of 9 curves, 
  is drawn for the larger values of  $c$ whereas the remaining  6 curves are for
  the smaller $c$'s. Note that although the difference in $c$ for any two  
  neighbouring curves is only 0.1 nevertheless the upper curves in the figure 
  differ significantly      in
  appearance from each other and from those of the central part.   
   \par
 Fendley and Tchernyshyov \cite{Fendley} have discussed one-dimensional phase
 transitions in systems with infinite number of degrees of freedom per site.
 They argue that a singularity of the maximum eigenvalue in the  "transfer
 matrix" (their model considers a set of systems with SU(N) type symmetry)
 causes a phase transition.  The $N \to \infty$ limit of
 \cite{Fendley} corresponding to an infinite number of degrees of freedom per
 site is replaced, in our case,  by a large (infinite) number   of barriers in a finite
 interval. The transfer matrix that we have introduced connecting neighbouring
 sites (barrier-gap structures) is independent of $\beta$ (inverse temperature),
  but the eigenvalue of the total transfer matrix of the infinite system has a
  branch point as we show in Section 3. This singularity can influence the
  behaviour of the partition function, resulting, as in \cite{Fendley}, in the
  phase transitions that we observe in our numerical study. Figure 3 shows 39
  different curves of the specific heat $C_h$ as function of the temperature for
  $N=15$. Each curve is for a different integral value of $c$ from the range 
  $2, 3, 4,\ldots 40$. As in Figure 1 one can see a dense batch of similar
  curves in the central part of the figure and other 16 curves that are graphed
  one above the other in the upper part of it. The dense batch corresponds, as 
  in Figure 1, to the larger values of $c$ and so may represent the solid
  materials that are characterized by a periodic structure (see the discussion
  of Figure 1). The other 16 curves correspond to the smaller values of $c$ but 
  compared to the former figure (for $N=6$) one can see that these curves
  demonstrate the double peak appearance found in 
 antiferromagnetic \cite{Tanaka} and superconducting materials \cite{Kim}. The
 first peak may be clearly seen  at low temperatures at about $T \approx 0.5$
 and the second higher peak at about $T \approx 5$. All the curves of Figure 3,  
 the Debye-like singly peaked as well as the doubly peaked curves,  merge together
 for large $T$ into one batch that tends to the value of $C_h(T>\!>1) \approx
 0.55$. Note that Figure 1 also shows the same general form of a central dense
 batch of Debye-like curves and other different graphs in the upper part of the
 figure. These graphs, however, show no sign of double peak, even when finely 
 graphed in the neighbourhood of the critical temperature. Thus, one may
 conclude that the double peak phenomenon is related to the number of barriers
 so that it is more apparent for the large number of them as seen from Figure 4
 which  
  shows 39 different curves of the specific heats, for $n=35$, as
functions of the temperature  in the range $0.1 \le T \le 100$. Each curve is 
for a different integral  
value of $c$ in the range  $2,3,\ldots,40$.  As in the former figures  the
 similar  Debye like curves in the central part of the figure  are for large 
 $c$'s and  the partition function, for these values of the parameter $c$, 
 therefore should
contain some mathematical features in common with the partition function for the
vibrational modes of solid crystals.      
  The other curves are for
 small $c$'s and they may correspond , as in the former figures, 
 to "soft"  
 substances. We note that seven of 
 the  curves   have each a part below the Debye-like curves and a part above 
 them and so they resemble, in a more apparent manner than Figure 3, 
  the remarked double peak phenomena 
  \cite{Tanaka,Kim}. 
 The second peak, in our case, is obtained at a comparatively large value of the
 temperature ($T\approx 40$) compared to the values ($T \approx 1$) in
 \cite{Leung,Tanaka,Kim} and to the value of $T \approx 5$ in Figure 3.  
  Note also that the maximum values obtained by the curves of this figure 
 are unity,  whereas, most of the curves in the former figures  have maxima that
 exceed unity.  All the curves of figure 4 show for small
 values of $T$, as in the former figures, 
    peaks that are suggestive of phase transitions. \par 
     Thus, as remarked, the
    double peak appearance is more pronounced for the larger values of $N$, 
     but we note that 
    as $N$  becomes larger the second peak diminishes in height until it
    disappears entirely for large enough $N$ and remains only the first peak for
    small $T$. But, as we see in Section 3, for the limit  $N \to \infty$ the
    double peak phase transition is seen  for a larger range of $c$ than
    for finite $N$.          
  \par 
We can find the corresponding critical exponent \cite{Reichl1} $\chi$ 
associated with these phase
transitions 
 by noting, after studying and analysing the behaviour of these curves in the
 immediate neighbourhood of the critical temperature $T_c$,  that we can write  
 an analytical approximate expression for the specific heat, in the
 neighbourhood of these points,  as follows
\cite{Reichl1} \begin{equation} \label{e11} C_h(\epsilon)=A+B\epsilon^{\frac{1}{2}}, 
\end{equation} where  $\epsilon=\frac{T-T_c}{T_c}$. As seen,  the first
order derivative of this specific heat with respect to the temperature diverges
at the point $T=T_c$ and so,     
the critical exponent $\chi$ is obtained as  \cite{Reichl1} 
\begin{equation} \label{e12} \chi=
1+\lim_{\epsilon \to
0}\frac{ln|\grave C_h(\epsilon)|}{ln(\epsilon)}=1+
\lim_{\epsilon \to
0}\frac{ln|\frac{B}{\epsilon^{\frac{1}{2}}}|}{ln|\epsilon|}=\frac{1}{2}, \end{equation}
where $\grave C_h(\epsilon)$ is the derivative of $C_h$ from Eq (\ref{e11}) with
respect to $\epsilon$ and  the unity value of the first term denotes the order (which is 1 here) of the
derivative of $C_h$ from Eq (\ref{e11}) which diverges at $T=T_c$, that is, 
the appropriate critical exponent is $\frac{1}{2}$. \par
The conspicuous departure of the curve of the specific heat,  for small $c$, 
from that of Debye-like behahiour  can be explained 
by noting that the total number of nondegenerate energies, as in \cite{Fendley},
 that satisfies Eq
(\ref{e6}) varies for different values of these $c$'s. Moreover,    we find, 
numerically, for all finite $N$ and for small $c$, no energy from the lower 
part of the spectrum   that
satisfies Eq (\ref{e6}) for the $e<v$ case. That is,   
solutions of Eq 
(\ref{e6})  for this case are found,  for small $c$'s,   only  
from the part of the spectrum that is close to the value of $v$. Thus, when we
sum upon all the allowed energies, in order to calculate the average energy and the
specific heat, the summation does not include   the lower part of the spectrum. 
 For example, the number of nondegenerate allowed (energies) solutions 
of Eq  (\ref{e6}),   for $N=6$,  $e<v$ and  $c=1.5$ are only 3, whereas they 
amount to 311 for  $c=15$. 
     That
is, for  the larger values of $c$  we have a larger number of additional (that 
may be  thousands for
large potential $v$) allowed energies,  and this 
yields entirely different values for the average energy $<e>$ and the specific
heat $C_h$ derived from it.  
  \par
  The entropy $S$ and the
specific heat $C_h$ are related by \cite{Reif} $C_h=T\frac{\partial S}{\partial T}$. 
Thus, from the last discussion 
we infer that also the change of the entropy $S$ with the temperature $T$ jumps
at the same values of $T$ in which the specific heat $C_h$ jumps.  That is, the
change of the entropy with $T$  has also phase transition. 
Moreover,  the entropy $S$ and the free 
energy $F$ are related by the equation \cite{Reif} $S=-\frac{\partial F}{\partial T}$, so
that, differentiating both sides of the last relation with respect to the
temperature $T$ and using the relation between the specific heat $C_h$ and the
entropy $S$ we obtain \begin{equation} \label{e13} C_h=T\frac{\partial S}
{\partial T}=-T\frac{\partial^2F}{\partial T^2} \end{equation} 
From the last equation we see that the second derivative of the free energy $F$
with respect to the temperature $T$ also changes steeply at the same values 
 of $T$ in which  $C_h$ does so. \par        
  When $c$ becomes very large we have $b>\!>a$, so that we may ignore $a$ compared
to $b$. Thus, writing the trigonometric functions of the components of $T$ from
Eq (\ref{e3}) as exponentials, substituting in Eq (\ref{e2}), and ignoring, as
noted, $a$ with respect to $b$ we can see that the  matrix $Q$
from Eq (\ref{e2}) becomes in the limit of a very large $c$ the two dimensional
unit matrix. In this case Eq (\ref{e6}), from which the energy spectrum is
obtained, becomes the same as Eq (\ref{e8}) from which the energy spectrum has
been 
obtained as $e=k^2=(\frac{\pi N}{C})^2$. But we note that whereas Eq (\ref{e8})
was obtained for the case of high energies only for which the index $n$
begins from a  large value $n_0$ ($n_0=941$), here, in the limit of very
large $c$,  $n$ assumes all integer values of $n=0, 1, 2, \ldots$. In this case the
average energy is \begin{equation} <e\!>_{C_{h_{ \to
\infty}}}(\beta)=\frac{\sum_{n=0}^{n=\infty} (\frac{\pi
n}{C})^2e^{-\beta (\frac{\pi n}{C})^2}}
{ \sum_{n=0}^{n=\infty} e^{-\beta (\frac{\pi n}{C})^2}} \label{e14} \end{equation} 
Note that $<e\!>_{C_{h_{ \to
\infty}}}(\beta)$  from the last equation does not depend upon 
the number of barriers $N$. The specific heat is \begin{equation} C_{h_{c \to
\infty}}=\frac{\partial <e>_{C_{h_{ \to \infty}}}(\beta)}{\partial T}=\frac{1}{T^2}(\frac{\sum_0^{\infty}
(\frac{\pi n}{C})^4e^{-\beta(\frac{\pi n}{C})^2}}
{\sum_0^{\infty}e^{-\beta(\frac{\pi n}{C})^2}}-\frac{\sum_{n=0}^{n=\infty}
(\frac{\pi n}{C})^2e^{-\beta(\frac{\pi n}{C})^2}
\sum_{\grave n=0}^{\grave n=\infty}
(\frac{\pi \grave n}{C})^2e^{-\beta(\frac{\pi \grave n}{C})^2}}
{\sum_{n=0}^{n=\infty}e^{-\beta(\frac{\pi n}{C})^2}
\sum_{\grave n=0}^{\grave n=\infty}e^{-\beta(\frac{\pi \grave n}{C})^2}})
\label{e15} \end{equation} 
Plotting  the curve of the specific heat from the last equation as a
function of the temperature (not shown here) one can  see that at 
small $T$ 
$C_{h_{\to\infty}}$ varies rapidly from zero to 0.52 from which it descends
sharply to its asymptotic value of 0.5. The curve is not differentiable at the
point at which it assumes the value of 0.52 and so this point appears to be  
 a phase transition one.   
 
 \bigskip \noindent 
\protect \section{The one-dimensional $N$ potential barrier system  for $N \to\infty$ \label{sec3}}
 We discuss, now, the case where the number of barriers $N$ tends to the limit  
 $N \to \infty$.  We may use  for this case   the equations 
 (\ref{e1})-(\ref{e2}) derived for the
 finite $N$ case in the previous section, so that    
   taking  the limit of a very
 large $N$ one  obtains  from Eq (\ref{e2}) for the right hand side of the
 potential barrier system at the point $x=\frac{a+b}{2}$ where  $n=N$
 \cite{Bar1} 
\begin{eqnarray} && \left[
\begin{array}{c} A_{2N+1} \\ B_{2N+1} 
\end{array} \right]=\left[ \begin{array}{c c} e^{-ik(a+b)}
 & 0 \\0& e^{ik(a+b)} \end{array} \right] \Biggl( \left[ \begin{array}{c c}
e^{\frac{ikb}{N}}&0\\
0&e^{-\frac{ikb}{N}} \end{array} \right] \left[ \begin{array}{c c}
T_{11}&T_{12}
\\T_{21}&T_{22}\end{array} \right] \Biggr) ^{N} \left[ \begin{array}{c} A_1 \\ B_1 
\end{array} \right] =\nonumber \\ && = \left[ \begin{array}{c c} e^{-ik(a+b)}
 & 0 \\0& e^{ik(a+b)} \end{array} \right](1+\frac{i}{N}((kb+aq\frac{\xi}{2})
 \sigma_3+iaq\frac{\eta}{2}\sigma_2))^N\left[ \begin{array}{c} A_1 \\ B_1 
\end{array} \right] = \label{e16} \\ &&=\exp(-ik(a+b)\sigma_3)\exp(i((kb+\frac{aq\xi}{2})\sigma_3+
\frac{iaq\eta}{2}\sigma_2)) \left[ \begin{array}{c} A_1 \\ B_1
\end{array} \right] \nonumber \end{eqnarray} 
The middle expression was  obtained by expanding in a Taylor series the cosine and 
sine functions and keeping  only terms of the order $\frac{1}{N}$ and  the
last result by using   the 
relation $\lim_{n\to\infty}(1+\frac{c}{n})^n=e^c$, where $c$ is some constant. 
The    $\sigma_2$, and 
$\sigma_3$ are the two dimensional Pauli matrices $ \sigma_2=
\left[ \begin{array}{c c} 0&-i\\i&0 \end{array} \right]$,  
$\sigma_3=\left[ \begin{array}{c c} 1&0\\0&-1 \end{array} \right]$. 
The second exponent  of the last result of 
Eq (\ref{e16}) may be expanded in a Taylor series, so that after collecting 
corresponding terms we obtain 
\begin{equation} \label{e17} e^{i((kb+aq\frac{\xi}{2})\sigma_3+
iaq\frac{\eta}{2}\sigma_2)}=\cos(\sqrt{f^2-d^2})+\frac{i(f\sigma_3+id\sigma_2)}
{\sqrt{f^2-d^2}}\sin(\sqrt{f^2-d^2}),  \end{equation} where $f$ and $d$ are 
defined as  \cite{Bar1} $$ f=kb+aq\frac{\xi}{2},  \; \; \; d=aq\frac{\eta}{2} $$
 Thus,  making use of the relation  
 $(f\sigma_3+id\sigma_2)^2=f^2-d^2=\phi^2$, and defining $z=k(a+b)$ we obtain
 from Eqs (\ref{e16})-(\ref{e17}) for the $e>v$ case  \cite{Bar1}
\begin{equation} \label{e18} \left[ \begin{array}{c} A_{2N+1} \\ B_{2N+1}
\end{array} \right]=\left[ \begin{array}{c c}
e^{-iz}(\cos{\phi}+if\frac{\sin(\phi)}{\phi}) &ie^{-iz}d\frac{\sin(\phi)}{\phi} 
\\ -e^{iz}d\frac{\sin(\phi)}{\phi}&e^{iz}(\cos{\phi}-if\frac{\sin(\phi)}{\phi})
\end{array} \right]\left[ \begin{array}{c} A_1 \\ B_1
\end{array} \right] \end{equation} 
 For the $e<v$ case we use Eqs (\ref{e4}),   and  the corresponding
 quantities \cite{Bar1} $\grave f$, $\grave d$ and $\grave \phi$ 
 $$\grave f=kb-\frac{aq\eta}{2},  \; \; \; 
\grave d=\frac{aq\xi}{2}, \; \; \; \grave \phi^2=(\grave f \sigma_3-i\grave d\sigma_2)^2=
\grave f^2-\grave d^2, $$ to obtain  the 
following matrix equation equivalent to Eq 
(\ref{e18}). \begin{equation} \label{e19} \left[ \begin{array}{c} A_{2N+1} \\ B_{2N+1}
\end{array} \right]=\left[ \begin{array}{c c}
e^{-iz}(\cos{\grave \phi}+\frac{i\grave f\sin(\grave \phi)}{\grave \phi}) 
&-ie^{-iz}\frac{\grave d\sin(\grave \phi)}{\grave \phi} 
\\ ie^{iz}\frac{\grave d\sin(\grave \phi)}{\grave \phi}&
e^{iz}(\cos{\grave \phi}-
i\frac{\grave f\sin(\grave \phi)}{\grave \phi})
\end{array} \right]\left[ \begin{array}{c} A_1 \\ B_1
\end{array} \right],  \end{equation} 
 We can, now, find the energy epectrum of the dense system in an equivalent way 
to the finite $N$ case of the previous section.  For both cases of 
$e>v$ and $e<v$,  we obtain  equations similar to 
Eq (\ref{e5}), but now the two dimensional matrices  $Q$ are those on
the right hand side of Eqs (\ref{e18}),(\ref{e19}), where their  
four components are given 
explicitly. Note that the four components of the two dimensional matrix $Q$ 
for  finite $N$ (see Eq (\ref{e2})) can  be obtained only numerically. 
  Thus, using, for the $e>v$ case,  the explicit expression of $Q$ from Eq
  (\ref{e18}) 
  we can write
  the analogous equation (for $N \to \infty$) to Eq (\ref{e6}) as 
\begin{eqnarray} && \det\left[ \begin{array}{c c} 
\frac{e^{2ikC}}{Q_{22}}-1& \frac{e^{2ikC}Q_{12}}{Q_{22}}\\
-\frac{e^{2ikC}Q_{21}}{Q_{22}}  & \frac{e^{2ikC}}{Q_{22}}-1
\end{array} \right]=\frac{e^{4ikC}}{Q^2_{22}}-\frac{2e^{2ikC}}{Q_{22}}+1+
\frac{Q_{12}Q_{21}e^{4ikC}}{Q^2_{22}}=\label{e20} \\ 
&&=\cos(4kC)(1+\frac{d^2\sin^2(\phi)}{\phi^2})+\cos(2z)(\cos^2(\phi)-
\frac{f^2\sin^2(\phi)}{\phi^2})+\frac{2f\sin(\phi)}{\phi}(\sin(2z)\cos(\phi)-
\nonumber \\ && -\sin(2kC+z))-2\cos(2kC+z)\cos(\phi)+ 
i(\sin(4kC)(1+\frac{d^2\sin^2(\phi)}{\phi^2})+\sin(2z)(\cos^2(\phi)-
\nonumber \\ &&-
\frac{f^2\sin^2(\phi)}{\phi^2}) +\frac{2f\sin(\phi)}{\phi}
(\cos(2kC+z)-\cos(2z)\cos(\phi))-2\sin(2kC+z)\cos(\phi)) =0 \nonumber 
\end{eqnarray}
For the $e<v$ case we use the explicit expression of $Q$ from Eq (\ref{e19}) 
  to 
obtain a similar equation to Eq (\ref{e20}) from which the energy spectrum for
the $e<v$ case may be obtained.  \par
Defining the parameters $\kappa$ and $\tau$ as $$e^{i\kappa}=\frac{\cos(\phi)+
i\frac{f\sin(\phi)}{\phi}}{\sqrt{\cos^2(\phi)+\frac{f^2\sin^2(\phi)}{\phi^2}}}, 
\ \ \ \  \tau=1+\frac{d^2\sin^2(\phi)}{\phi^2}, $$ one can obtain the eigenvalues of
either Eq (\ref{e18}) or (\ref{e19}) in the form 
\begin{equation} \label{e21} \lambda_{1,2}=\tau \cos(\phi-\kappa) 
\pm\sqrt{\tau^2 \cos^2(\phi-\kappa)-1} \end{equation} The eigenvalues of the
$e>v$ case are obtained by substituting the correct $\phi$ (see the inline
equation after Eq (\ref{e17}))  and those 
of the $e<v$ case by substituting the corresponding quantity (see the displayed
equation after Eq (\ref{e18})).   
From  Eq (\ref{e21}) one can see that the  derivatives of the eigenvalues 
$\lambda_{1,2}$ may be singular at certain values of $\phi$ so that they fulfil
the condition in \cite{Fendley} for finding phase transition in a one
dimensional system. This condition is necessary but may not be sufficient. Our
numerical results suggest that such a transition occurs.  \par
The energy spectrum is composed from those energies that satisfy the real and
imaginary parts of the last equation for the $e>v$ case and the corresponding
one for the $e<v$ case. Thus,  we may obtain the
average energy $<\!e\!>_{N\to \infty}(c)$ for each value of $c$ as in Eq
(\ref{e9}) (without, of course the dependence upon $N$).
From these average energies we obtain the corresponding specific heats
$C_h(c)_{N\to \infty}$ 
as functions  of $c$ and $T$. Note that although Eq (\ref{e20}), from which we
derive the
average energy $<\!e\!>_{N\to \infty}(c)$ and the specific heat 
$C_h(c)_{N\to \infty}$, is obtained analytically compared to the corresponding
Eq (\ref{e6}) for finite $N$, nevertheless, these $C_h(c)_{N\to \infty}$ have
the same form as those  of the finite $N$ (obtained by differentiating Eq
(\ref{e9}) with respect to $T$) and are therefore difficult to study
analytically. This is true, especially, for small $T$ where the phase
transitions are generally encountered. \par
The dependence of $C_h(c)_{N\to \infty}$ upon the temperature 
$T$, as a 
function of $c$,    is, for small $T$,  different from  
the dependence   discussed in the previous section for finite $N$. That is, 
it jumps up to its peak value from which it immediately jumps down to rise 
again to anther higher maximum. 
We note that, generally,  for finite $N$ there is only one peak maximum, and
although in Figures 3 and 4 we see  several curves that have double peaks,
nevertheless,  this is only for small $c$ and that when $c$ grows
the curves become the same as that of Debye as seen in the dense central part
of Figures 3-4 which are for large $c$. Compared to this the double peak
appearance of the specific heat curves for infinite $N$ is retained even for
large $c$ as can be seen from Figure 5 which is drawn for $c=200$. Figure 6 
shows 30 
different curves of the specific heats as function of the temperature. Each 
curve is for a different   value of $c$ from the set $0.3,0.4,\ldots,3.2$. Note the 
large difference in the heights of the two peaks of each curve, and that both 
are points 
 where the first derivative  with respect to the
 temperature $T$ attains a very large value  and so they appear to be 
   phase transition points. A similar discussion to that of the finite $N$
 case (see also Eqs (\ref{e11}), (\ref{e12})) yields a critical exponent of
 $\frac{1}{2}$  
 for both peaks.  That is, producing these curves in a fine grained manner in
 the immediate neighbourhood of the critical temperature $T_c$ we realize, as
 for finite $N$ (see the discussion before Eq (\ref{e11})) that we may
 approximate analytically the form of $C_h(c)_{N\to \infty}$ by Eq (\ref{e11})
 with a critical exponent of $\chi \approx 0.5$. Indeed, comparing the forms of
 the curves for finite $N$ to those of the infinite $N$ in the neighbourhood 
 of the critical temperature $T_c$ one does not find a large differece.
 Moreover, as $c$ grows both kinds of curves show generally, except for the
 double peak appearance which is retained for the infinite $N$ even for large
 $c$, the same behaviour which characterizes the Debye-type curves. This may be
 seen from Figure 7 which shows 40 different curves of the specific heat as
 functions of the temperature $T$ for integral values of $c$ from $2, 3, 4,
 \ldots 41$. Comparing these curves to those of Figure 6 which are drawn for
 smaller $c$ one sees that the dense batch of similar curves in the central  part
 of the figure,  which characterizes the solid crystals (as we have 
 encountered in
 the former figures),  have appeared also for the infinite $N$ case. This is
 because of the larger values of $c$, for which Figure 7 is graphed, that
 enable, as for the finite $N$ case, a Debye-like forms for these curves. Note
 that this form is absent in Figure 6 (and also in Figure 2) 
  because all the curves there are drawn for
 small $c$. The other curves in Figure 7 that are not part of the central dense
 batch are for the smaller values of $c$ and, therefore, they are similar to
 those of Figure 6. As noted, the difference between the finite and infinite $N$
 lies, especially, in the double peak phenomenon that is seen in the infinite
 $N$ case even for large values of $c$ as can be seen from Figure 5. When
 Figure 7 is produced in a fine grained manner in the neighbourhood of the
 critical temperature $T_c$ one can see clearly the double peak for any curve as
 in Figure 5.    As the temperature increases all the
curves tend to the value of 0.55 as for the finite $N$ case. When $c$ becomes
very large 
the curves (not shown here) of the specific heats become similar to each other
and to the Debye graph.  \par  
We infer from the former results   that also the first derivative  of the 
entropy $S$ and the second
derivative of the free energy $F$, both with respect to the temperature,  change
in an abrupt manner  at the same values  of $T$ (see the analogous 
discussion at the previous section). 
\bigskip \noindent 
\protect \section{Concluding Remarks  \label{sec5}} \smallskip 
We have shown that the one-dimensional multibarrier potential of finite range
shows signs of phase transitions in specific heat for certain values of the temperature $T$.
These phase transitions depend upon the number of barriers $N$ and the ratio
$c$ of the total spacing to their total width and are demonstrated for 
  both cases of  finite and infinite  number $N$ as shown in figures 1-7 and
  also for small and large values of $c$. Moreover, it is seen from the curves
  of the specific heat as a  function of the temperature for $N=35$ and small
  $c$ (see Figure 4) and for infinite $N$ and a large range of  $c$ 
  (see Figures 5-7) 
  that the phase transitions appear in a double peak form. Double peaks have
  been seen in antiferromagnetic and superconducting materials and are
  apparently associated with dynamically induced phase transitions
  \cite{Leung,Tanaka} and in the
  quark-gluon plasma \cite{Ko}. \par 
  We note that we have found \cite{Bar1} that the one-dimensional multibarrier
  system discussed here demonstrates also, for large $N$, a unit value for the
  transmission probability and signs of chaos which may be interpreted in terms
  of effective decoherence and  the space analog \cite{Bar2} of the Zeno 
  effect \cite{Zeno} in which a very large number
  of repetitions of the same experiment (interaction), in a finite total time, 
   preserves the initial
  state of the system. It has also been shown  \cite{Peres,Bar2} that a
  beam of light that passes through a large number of analyzers arrayed along a
  finite interval of a spatial axis, a configuration which is very similar to
  the one discussed here,  remains after the passage  with the same initial 
  polarization and
  intensity it had before passing. This kind of preservation of the initial
  ``state'' by passing through a large number of physical apparatuses,  each of
  them is supposed by itself to change the state of the passing system, has also
  been shown in the {\it classical}  regime \cite{Bar3} where the initial density of
  classical particles passing through a one dimensional array of imperfect traps
  \cite{Smol} remains at the same value it had before the passage if the ratio
  of the total spacing to width (which corresponds to the ratio $c$ here)
  increases.  \par
   Thus, our finding here that when $c$ grows the curves of the
  specific heat, as functions of the temperature $T$, become similar to the
  known graph of Debye \cite{Reif},  indicates that the system makes a
  transition  
   to the
  physical situation which behaves like the vibrational modes of  a  solid 
  crystal.   \par
  We have found that the critical exponents associated with these phase
  transitions have the value $\frac{1}{2}$.  The other
  statistical parameters associated with  the specific heat such as the entropy
  and the free energy also demonstrate, in the rate of their changes with
  respect to the temperature $T$, the same type of behaviour at the same
  values of $N$, $c$, and $T$.

    \bigskip \bibliographystyle{plain}

\begin{thebibliography}{99}
\bigskip \parindent 1 in 
 
\bibitem{Bar1}       D. Bar and L. P. Horwitz, Eur. Phys. J.  B, {\bf 25}, 
                     505-518, (2002);  Phys. Lett A, {\bf 296}, 265-271, (2002).
\bibitem{Merzbacher} {\rm "Quantum mechanics"} $2^{nd}$ edition by E.Merzbacher, John 
                      Wiley and sons,  (1961); {\rm "Quantum mechanics"} by C. C. Tannoudji, B. Diu, And 
                       Franck Laloe, John Wiley and Sons (1977)        
\bibitem{Reichl} {\rm "The transition to chaos in conservative classical systems:
                   Quantum manifestations"},  L. E. Reichel, Springer,
		   Berlin, 1992 ;E. Haller, H. Koppel and L. S. Cederbaum, 
		   Chem. Phys. Lett, {\it 101}, 215-220, (1983); T. A. Brody, J.
		   Flores, J. B. French, P. A. Mello, A. Pandey and  S. S. M. Wong,  
		    Rev. Mod. Phys, {\bf 53}, 385, (1981)
 \bibitem{Reif}       {\rm "Statistical Physics"} by F. Reif,  McGraw-Hill book 
                      company, 1965;    {\rm "Statistical Physics"} by L. D.
		      Landau and E. M. Lifshits, Oxford, Pergamon Press, (1980).
\bibitem{Fendley}      P. Fendley and O. Tchernyshyov, ArXiv cond-mat/0202129		      
\bibitem{Leung}      K. -T.  Lueng and Z. Neda, Phys. Lett.  A, {\bf 246}, 505,
                     (1998).  
 \bibitem{Tanaka}  Y. Tanaka, H. Tanaka, T. Ono, A. Oosawa, K. Morishita, K.
                   Iio,  T. Kato, H. A. Katori, M. I. Bartashevich and  T. Goto, 
		   J. Phys. Soc. JPN, {\bf 70}, 3068, (2001); P. G. Pagliuso, R.
		   Movshovich, A. D. Bianchi, M. Nicklas, N. O. Moreno, J. D.
		   Thompson, M. F. Hundley, J. L. Sarrao, and Z. Fisk, 
		    arXiv: Cond-Mat/0107266, v2, 2001; 
\bibitem{Kim}       B. J. Kim, P. Minnihagen, H. J. Kim, M. Y. Choi and G. S.
                     Jeon, Europhys. Lett, {\bf 56}, 222, (2001). 
\bibitem{Ko}        C. M. Ko and M. Asakawa, Nucl. Phys A, {\bf 566}, 447c-458c,
                    (1994).  
\bibitem{Reichl1}  {\rm "A modern course in Statistical physics"} by L. E. Reichl,
                    University of Texas Press, Austin, (1980).		     		   
\bibitem{Yu}          K. W. Yu,  Computers in Physics,  {\bf 4}, 176-178, (1990) 
\bibitem{Flowers}    {\rm "Properties of Matter"} by B. H. Flowers and E.
                      Mendoza, John Wiley \& Sons Ltd. London, (1970). 
\bibitem{Maple}       {\rm ``Quantum Mechanics using Maple''}, by H. Marko,
                     Springer, Berlin, (1995). 		      
\bibitem{Zeno}  B. Misra and E. C. Sudarshan, J. Math. Phys,{\bf 18}, 756, 
                (1977);  
            {\rm "Decoherence  and the appearance of a classical world in 
            quantum theory"},   D. Giulini, E. Joos,  C. Kiefer, J. Kusch,
            I. O. Stamatescu and H. D. Zeh, Springer-Verlag, (1996); 
	    Marcus Simonius,  
            Phys. Rev. Lett, {\bf 40},  15, 980-983,  (1978);  R. A. Harris and 
	    L. Stodolsky, J. Chem. Phys, 
            {\bf 74}, 4, 2145,  (1981);  Mordechai Bixon, Chem. Phys, 
           {\bf 70}, 199-206 (1982);  Saverio Pascazio and Mikio Namiki, 
	    Phys. Rev A 
            {\bf 50},  6, 4582,  (1994); W. M. Itano, D. J. Heinzen, J. J. Bollinger, 
	    and 
            D. J. Wineland, Phys. Rev A {\bf 41}, 2295-2300,  (1990); R. J. Cook, 
	    Physica 
            Scripta T {\bf 21}, 49-51 (1988);  A. Peres, Phys. Rev D {\bf 39},  10, 
	    2943,  
            (1989); A. Peres and Amiram Ron, Phys. Rev A  {\bf 42},  9,  5720,  
	    (1990); 
	   Y. Aharonov and M. Vardi, Phys. Rev D,  {\bf 21}, 2235,  (1980); 
           P. Facchi,  A. G. Klein, S. Pascazio and L. Schulman,  Phys. Lett A 
            {\bf 257}, 232-240, (1999).
\bibitem{Peres}   A. Peres, Am. J. Phys, {\bf 48}, 931-932, (1980). 	    
\bibitem{Bar2}     D. Bar and L. P. Horwitz, Int. J. Theor. Phys, {\bf 40}, 
                    1697-1713,  (2001)	    
\bibitem{Bar3}      D. Bar,   Phys. Rev.  E, {\bf 64}, No: 2, 026108/1-10, 
                    (2001) 
\bibitem{Smol}     R. V. Smoluchowski, Z. Phys. Chem., Stoechiom.
                  Verwandtschaftsl, 29, 129, (1917); \\ {\rm "Diffusion and 
		  reactions in fractals and disordered
                   media"}
                   by D. Ben-Avraham And S. Havlin, Cambridge, 
		   Camgridge University Press, 2000;		       

  \end{thebibliography}

  \newpage

 \begin{figure}[hb]
\centerline{
\epsfxsize=3.5in 
\epsffile{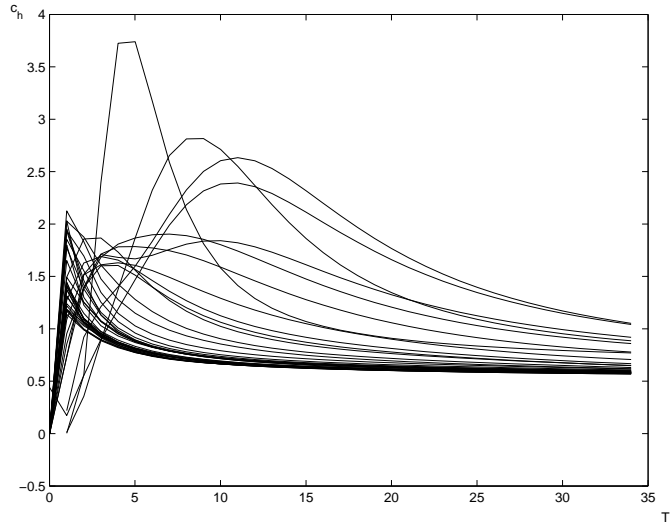}} 
 \caption[fig1]{38 different curves of the specific heat as a function of the
 temperature for $N=6$ and the following integral values of $c=2,3 \ldots 39$. 
 The dense
 central part of the graph are the curves obtained for the larger values of 
  $c$ and they all
 resemble, except for the sharp peaks at the left,  the Debye's graph for 
 solid crystals. The curves have very large values for the derivative 
  at the sharp peaks and so may be suggestive of 
 phase transition at these points. The other curves are for small
 $c$ and they represent the ``soft'' substances that have a comparable low
 Einstein frequencies and therefore a high values of the specific heats.  }
\end{figure}

 \begin{figure}[hb]
\centerline{
\epsfxsize=3.5in 
\epsffile{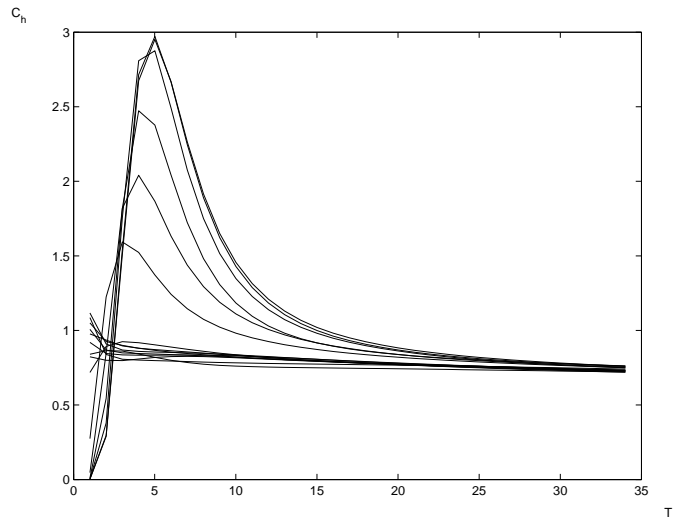}} 
 \caption[fig1]{15 different curves of the specific heat as a function of the
 temperature for $N=6$ and the following  values of $c=0.3, 0.4,  \ldots 1.7$. 
 The dense
 central part of the graph is composed  of 9 similar 
  curves obtained for the larger 
 values of 
  $c$ but since these values are small compared to those of Figure 1 these
  curves are not similar to the Debye graph. 
     The other 6 curves are for the smaller values of $c$ and they deviate 
 significantly  from each other and from those at the center. 
 Note that  a 
  small difference of only 0.1 in  $c$ between neighbouring curves  is capable of
  producing this large difference in appearance. Comparing 
   this figure with the former one which has been drawn for the same $N$ but 
   for
    larger values of $c$ one  sees that only 3 curves out of 38 in Figure 1
   deviate markedly from the dense central part. }  
\end{figure}

\begin{figure}[hb]
\centerline{
\epsfxsize=3.5in 
\epsffile{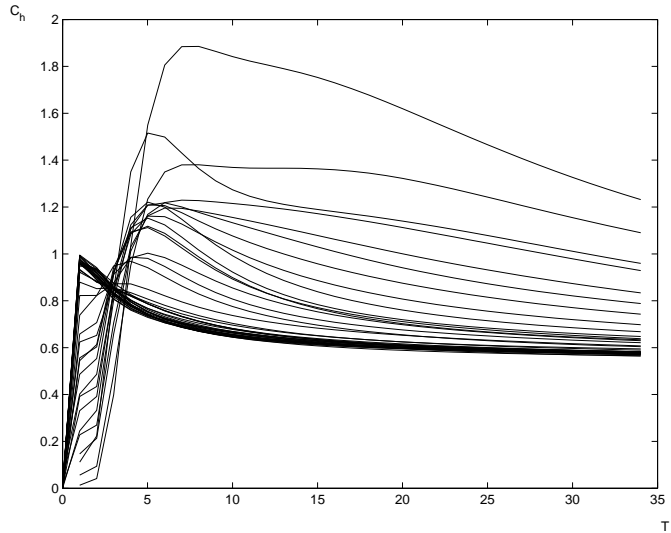}} 
 \caption[fig1]{39 different curves of the specific heat as a function of the
 temperature for $N=15$ and for the following integral values of $c=2,3 \ldots 40$. 
 As in
 figure 1 the dense central part of the figure are the curves that represent 
 the Debye curve, 
 except for the sharp peaks at the left at which the specific heats have  very
 large values for their derivatives 
  and so are  suggestive of phase transition. Note that the curves that 
  are not part
  of the central dense batch have double peaks where the first one may be seen at
   low $T$.}
\end{figure}

\begin{figure}[hb]
\centerline{
\epsfxsize=3.5in 
\epsffile{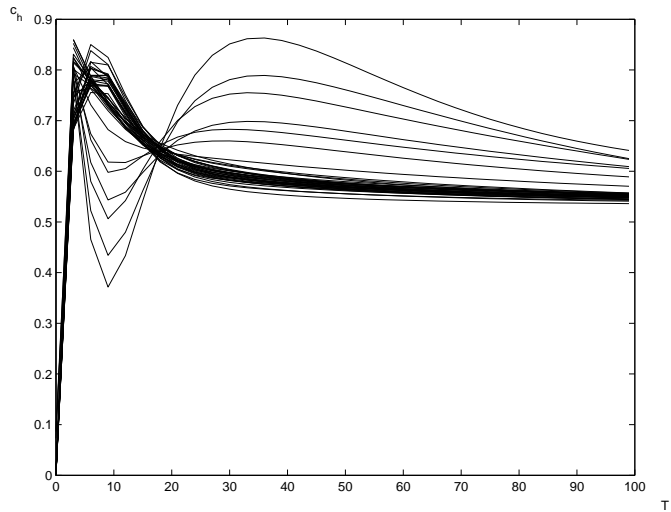}} 
 \caption[fig1]{39 different curves of the specific heat as a function of the
 temperature for $N=35$ and for the following integral values of $c=2,3 \ldots 40$. 
 As in the former figures 
  the dense central part of the figure are the curves that represent 
 the Debye curve, 
 except for the sharp peaks at the left at which the specific heats have very
 large values for the derivatives. 
 Note that the double peak character of the curves that are separate from the
 central part is more pronounced than at the former figure.}
\end{figure}

  \begin{figure}[hb]
\centerline{
\epsfxsize=3.5in 
\epsffile{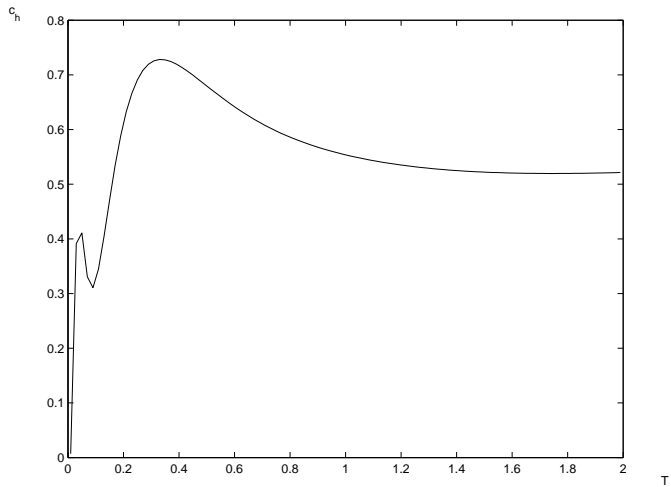}} 
 \caption[fig1]{Double peak  of the specific heat as a function
 of the temperature for infinite $N$ and for $c=200$. Note the  difference
 between the heights of the two maxima and that the double peak is retained even
 at this large value of $c$ compared to the finite $N$ case.}
\end{figure}

\begin{figure}[hb]
\centerline{
\epsfxsize=3.5in 
\epsffile{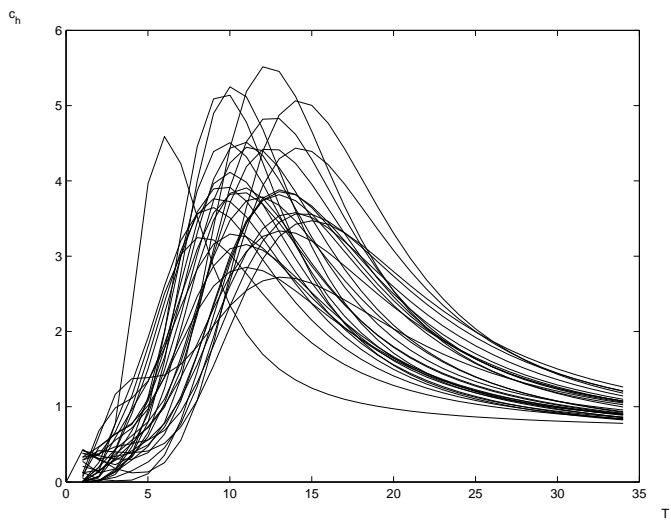}} 
 \caption[fig1]{30 different curves of the specific heat as a function of the
 temperature for infinite $N$ ans for the following values of $c=0.3, 0.4, \ldots
 3.2$. All the curves  tend to the value of 0.55 for large $T$. The first peaks 
 are not clearly shown for all the curves due to the large range of $T$ over 
 which they are drawn. For smaller ranges of $T$, as in Figure 5, all the
 first peaks are clearly shown (not here).  Note that the
 second peak of each curve is much larger than the first and that all these
 second peaks are obtained for $T>5$. The peaks of each curve have very much
 large values for their 
 derivative  and so they are suggestive of phase transitions.  }
\end{figure}

 \begin{figure}[hb]
\centerline{
\epsfxsize=3.5in 
\epsffile{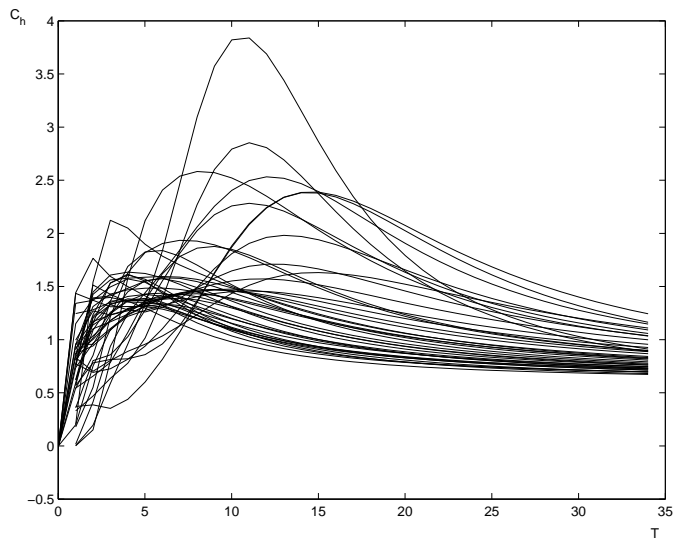}} 
 \caption[fig1]{40 different curves of the specific heat as a function of the
 temperature for infinite $N$ ans for the following integral 
 values of $c=2, 3,\ldots
 41$. Note the central dense batch of the similar figures known from finite $N$.
  It appears due to   the  large values of $c$ compared 
 to these for which Figure 6 (and also Figure 2 for the finite $N$) 
  was drawn which is the reason 
  that the  Debye-like form is absent in All 
 the curves  of Figure 6 (and Figure 2). Producing the curves of 
 Figures 6 and 7 in the
 immediate neighbourhood of the crtical temperature $T_c$ demonstrate clearly 
 the
 existence of their first peaks (not shown here).      }
\end{figure}
    
\end{document}